\documentstyle[preprint,aps]{revtex}

\begin{document}
\draft
\title{Stability of vortex lines in liquid
 $^3$He-$^4$He mixtures at zero temperature}
\author{Dora M. Jezek$^1$, Montserrat Guilleumas$^2$,
Mart\'{\i} Pi$^3$, and Manuel Barranco$^3$.}
\address{$^1$Departamento de F\'{\i}sica, Facultad de Ciencias Exactas
y Naturales, \\
Universidad de Buenos Aires, RA-1428 Buenos Aires, Argentine}
\address{$^2$Dipartimento di Fisica, Universit\`a di Trento. 38050
Povo, Italy}
\address{$^3$Departament d'Estructura i Constituents de la Mat\`eria,
Facultat de F\'{\i}sica, \\
Universitat de Barcelona, E-08028 Barcelona, Spain}

\maketitle

\begin{abstract}

At low temperatures   and  $^3$He concentrations   below $\sim 6.6 \% $,
 there is experimental evidence about the existence in liquid helium
mixtures, of stable vortices with  $^3$He-rich cores.
When the system is either supersaturated or submitted to a tensile
strength, vortices lose stability
becoming metastable and eventually completely unstable,
 so that their  cores freely expand.
 Within a density functional approach, we have determined the
 pressure-$^3$He concentration curve along which this  instability
appears at zero temperature.

\end{abstract}


The structure, stability and dynamics of
quantized vortices in
  superfluid $^4$He have been extensively   studied
 either experimental or theoretically (for a systematic review
 see \cite{don1}, and for recent work, see for example
\cite{da1,ma1,ni1,sa1,or1} and Refs. therein). However,
lesser work has been done in the case of $^3$He-$^4$He solutions
\cite{don1}.
An interesting aspect of the vortex structure in these mixtures was
disclosed by Williams and Packard \cite{wi1}, who provided experimental
evidence that, in diluted $^3$He-$^4$He solutions at low temperature (T),
stable vortices present  $^3$He condensation  onto the  core.

In this work  we address the problem of the stability
of a vortex line as a function of pressure (P) and $^3$He
concentration ($x$).
In particular, we will discuss the implications that the existence of
$^3$He-rich vortices may have on the critical supersaturation of
isotopic helium mixtures, and will determine the T$=$0 vortex
spinodal line as a function of $x$.

The interest in studying the stability of vortex lines at
negative pressures stems from recent attempts to describe the
phenomenon of quantum cavitation in liquid $^4$He \cite{ma1}.
In particular, it has been found  \cite{da1} that at T$=$0,
vortex lines become unstable at    P $\sim$ -8 bar, whereas  the
spinodal point is at P$\sim$ -9.4 bar \cite{bo1}, thus quantitatively
showing that vorticity raises the spinodal line (see also \cite{ma1}).

In $^3$He-$^4$He mixtures the situation is more complex. Indeed, a
vortex may become unstable either increasing the $^3$He
concentration,  or submitting the solution to a tensile strength that
may originate a negative pressure. A characteristic of the zero
temperature P-$x$ phase diagram that makes more intrincate the study
of vortex lines in this system,  is the existence of a demixing
line P$_d$($x$), experimentally determined from saturation up to 20
bar \cite{eb1}. A  recent calculation \cite{gui1} has found
that line to continue down to $x\sim$ 2.4 \%
and P$\sim$ -3.1 bar, which is the T$=$0 spinodal point of pure
$^3$He.  The existence in the
negative pressure, metastable region, of this equilibrium line between
pure $^3$He and the mixture, affects
the description of cavitation caused either by bubble growing
\cite{gui1,gui2} or by vortex destabilization.

Instabilities caused by supersaturation were studied in  \cite{je1}
within the so-called Hollow Core Model (HCM), adapted to helium
mixtures  replacing the  hollow by a $^3$He-rich core.
 This study was carried out
at positive pressures, where the model works better.
We have predicted that at P$=$0,  vortices become completely
 unstable for $x>$8.2\%.

Although HCM might seem  too  crude a model, at positive
pressures and close to the demixing line it bears the
basic physical ingredients, so it can be used as an useful
guide to understand  the appearance of unstable vortices.
Let us just recall that within HCM,
 the total energy per unit vortex
length as a function of the core radius R reads \cite{je1}:
\begin{equation}
\Omega_{HCM} =  S R + V R^2 + E_0 ln(R_\infty /R) \,\, .
\label{eq1}
\end{equation}
 It is the sum of an S-"surface term",
  plus a V-"volume term", plus a kinetic energy term.
 When the vortex is in the stable phase, all
three terms are positive, and only  stable vortices may exist.

On the contrary, in the metastable phase
the factor V in the volume term  becomes negative.
Depending on whether the system is underpressured or
 supersaturated, this factor is proportional either to the
 difference $\Delta$P between the  pressure in the hollow and in the
bulk,  or to the  difference  $\Delta\mu$
between the chemical potential of $^3$He in the mixture and of
pure $^3$He \cite{gui2}.
 As a consequence, the stable vortex becomes metastable
 and there exists a  critical vortex configuration for which the
 potential barrier has a maximum  located at a core  radius R$_c$.
 At the saddle point, when the energy difference between critical
and metastable vortices vanishes, the vortex core freely expands.

These  arguments can be made quantitative within the
density functional approach. To this end, we resort to the density
functional of Refs. \cite{gui1,gui2} which describes the
basic thermodynamical properties of liquid helium
mixtures at zero temperature.
For the sake of simplicity, we address the problem of a
vortex line. Using cylindrical coordinates  and taking the vortex line
as z-axis, the density profiles depend on the r-distance  to the z-axis
 and are obtained  solving the Euler-Lagrange equations
\begin{equation}
\frac{\delta\omega(\rho_3,\rho_4)} {\delta\rho_q} = 0 \,,\,  q=3,4,
\label{eq4}
\end{equation}
where $\omega(\rho_3,\rho_4)$ is the  grand potential density
functional \cite{gui1,gui2} to which we have added a centrifugal term
 $ \hbar^2 \rho_4 n^2 / (2 m_4 r^2) $ \cite{da1} associated with
the superfluid flow. We choose the quantum circulation number
n=1 because it corresponds
to the most stable vortex  \cite{da2}, $ m_4 $ is the $^4$He
atomic mass,
and $\rho_q$ are the particle densities of each helium isotope.

For  given P and $x$, Eqs. (\ref{eq4}) are solved imposing that at
long distances from the z-axis, $\rho_q$ equals that of the
metastable, homogeneous liquid $\rho_q^h$ ($x$ is simply
$\rho_3^h/(\rho_3^h+\rho_4^h$)), and that $\rho_4$ and the r-derivative
of $\rho_3$ are zero on the z-axis.  Notice that
the metastable and critical configurations
 are solutions of these equations for the {\em same}  P
and $x$ conditions. The barrier height  per unit vortex length is:
\begin{equation}
\Delta\Omega =2 \pi \int  r {\rm d}{r}
\left[\omega(\rho_3^c,\rho_4^c)-\omega(\rho_3^m,\rho_4^m)\right]\,
,
\label{eq5}
\end{equation}
where $\rho_q^c$  and $\rho_q^m$ are the particle densities of
the critical and metastable vortices, respectively. It is worth to
note that $\Delta\Omega$ is a {\em finite} quantity:
there is no need to introduce any r-cutoff as it would have been
 unavoidable if we had described either configuration separately.

Depending on the situation of the metastable vortex in the P-$x$
plane, there may  exist two  different kinds of critical
configurations. To illustrate it, we show in Fig. 1 the P-$x$ phase
diagram at T$=$0 \cite{gui1} and three selected metastable
configurations labeled 1 to 3. The grey zone represents the
stable region, and the dashed line is the demixing line.
Configuration 1 is underpressed, configuration 2 is
supersaturated and configuration 3 is both.
In all three cases, the  metastable configuration
corresponds to a rather compact vortex filled with $^3$He whose
radius increases with increasing $x$. This is not the case for the
critical vortex. Indeed, as configuration 2 is in the
supersaturated region, the critical vortex may have a large R$_c$ radius
if that point is close  enough to the demixing line, and
its core is filled with almost pure $^3$He. R$_c$
diverges at the demixing line, and also $\Delta\Omega$.   Since
configuration 1 is underpressed and undersaturated, the
critical vortex  also has a large radius provided point 1 is close
to the P$=$0 line, but its core is almost empty, with the surface
covered by $^3$He (Andreev states). R$_c$
diverges at the P$=$0 line, and also $\Delta\Omega$.
 As configuration 3 is
underpressed and supersaturated, it has two possible
critical configurations, one bearing the  characteristics of
configuration 1, and another bearing those of configuration 2.

These possibilities are displayed in Fig. 2. Fig. 2 (a)
corresponds to a type 1 configuration with P$=$-1.66 bar, $x=$1\%,
 whereas
Fig. 2 (b) corresponds to a type 2 configuration with P$=$0.91 bar,
$x=$8\%.

Fig. 3 shows the barrier height per unit length as a function of
P for $x=$1 to 9\%. For the sake of illustration, we
display for $x=$4\%, the barriers corresponding to the two
kinds of critical vortices already discussed: the dashed (solid)
 line is $\Delta\Omega$ for  empty- (filled-) core configurations.
 Notice that these curves have different
  slopes because they diverge at different pressures, the former at
P$=$0, and the later at P$=$P$_d$.
  For a given $x$, the P-value at which
$\Delta\Omega$ is negligible defines a point along the vortex
spinodal curve. That curve is the solid line in Fig. 1.

Fig. 4 shows the core radius of the saddle configurations as a
function of $x$ (solid line). Following \cite{or1}, we have defined that
radius as the r-value at which the superfluid circulation current
$\rho_4^c$(r)/r has a maximum.
We have found that  metastable vortices in the mixture
have a core radius  larger than in pure $^4$He \cite{da1}. This is in
agreement with  the experimental findings for
stable vortices \cite{don1}. Also shown in that figure is the radius of
the stable vortex at P$=$0 (dashed line), which is actually
metastable above $x=$6.6\%.

The above results have implications on the critical supersaturation
degree $\Delta x_{cr}$ of isotopic helium solutions at low temperatures.
Recent experiments \cite{sat1,mi1} have found $\Delta x_{cr}$ below
$\sim$1\%, whereas classical nucleation theory yields $\sim$10\%
\cite{je1,li1}. The microscopically calculated spinodal line \cite{kr1}
is about the same value. Within the HCM, we have argued \cite{je1}
that the rather small degree of supersaturation experimentally found
could be due to destabilization of $^3$He-rich vortices. The
density functional approach yields $\Delta x_{cr}$ values around
2\% (we recall that the maximum solubility at P$=$0 is
$\sim$ 6.6 \% \cite{la1}). This is shown in Fig. 5 for P$=$0, 0.5 and 1
bar. A discrepancy with experiment still exists. It is unclear whether
considering more realistic vortex geometries, like vortex rings, could
bring theory closer to experiment. Other possibilities to improve
on the agreement, such as vortex
destabilization due to quantum tunneling through,  or to
thermal activation over the barrier, seem to be ruled out. The former
because of the extremelly small quantum-to-thermal crossover temperature
\cite{bu1}, and the later because of the large mass of the critical
vortex \cite{je1}.

Vortex destabilization at negative pressures may also have implications
on the phenomenon of cavitation in isotopic helium solutions
 \cite{gui1,gui2}. Realistic calculations \cite{je2} indicate that
for a $^3$He concentration as small as $x\sim$1\%,
 cavitation driven either by quantum or by thermal
fluctuations triggers phase separation if the system is submitted
 to a tensile strength of 8.2 bar. If there are metastable
vortices in the mixture, the present calculations show that for much
smaller tensile strengths, of about 5.2 bar, the solution undergoes
phase separation. A delicate question is whether $^3$He atoms have
enough time to diffuse into the vortex core on the time scale of
current cavitation experiments \cite{ba1} when their concentration
is too small.

This work has been supported in part by the CICYT,  by the
 Generalitat de Catalunya Visiting Professors  and
GRQ94-1022 programs, and by the CONICET (Argentine) Grant No. PID 97/93.

\begin{figure}
\caption{ P-$x$ phase diagram at T$=$0. The grey zone represents
the stable region. The dashed curve is the demixing line,
which ends at the P$=$-3.12 bar, $x=$ 2.43\% cross,  and
the solid curve is the vortex spinodal line.}
\label{fig1}
\end{figure}
\begin{figure}
\caption{ Panel (a): vortex profiles for $x=$1\% and
P$=$-1.66 bar. Panel (b): vortex profiles for $x=$ 8\% and P$=$0.91 bar.
The solid lines  represent the total particle density, and the
dash-dotted (dashed) lines,  the $\rho_4$ ($\rho_3$) densities.
Critical (metastable) configurations are denoted as $\rho^c$
($\rho^m$). }
\label{fig2}
\end{figure}
\begin{figure}
\caption{ Barrier height per unit vortex length as a function of P for
the indicated $^3$He concentrations. At $x=$ 4\%, the dashed line
corresponds to empty-core configurations, whereas the solid line
corresponds to filled-core ones.}
\label{fig3}
\end{figure}
\begin{figure}
\caption{ Solid line, radius of the saddle vortex core.
 Dashed line, radius of the stable vortex at P$=$0.}
\label{fig4}
\end{figure}
\begin{figure}
\caption{ Barrier height per unit vortex length as a function of $x$
for P$=$0, 0.5 and 1 bar. The corresponding critical $x$-values are
8.46, 8.74 and 8.97\%, respectively.}
\label{fig5}
\end{figure}
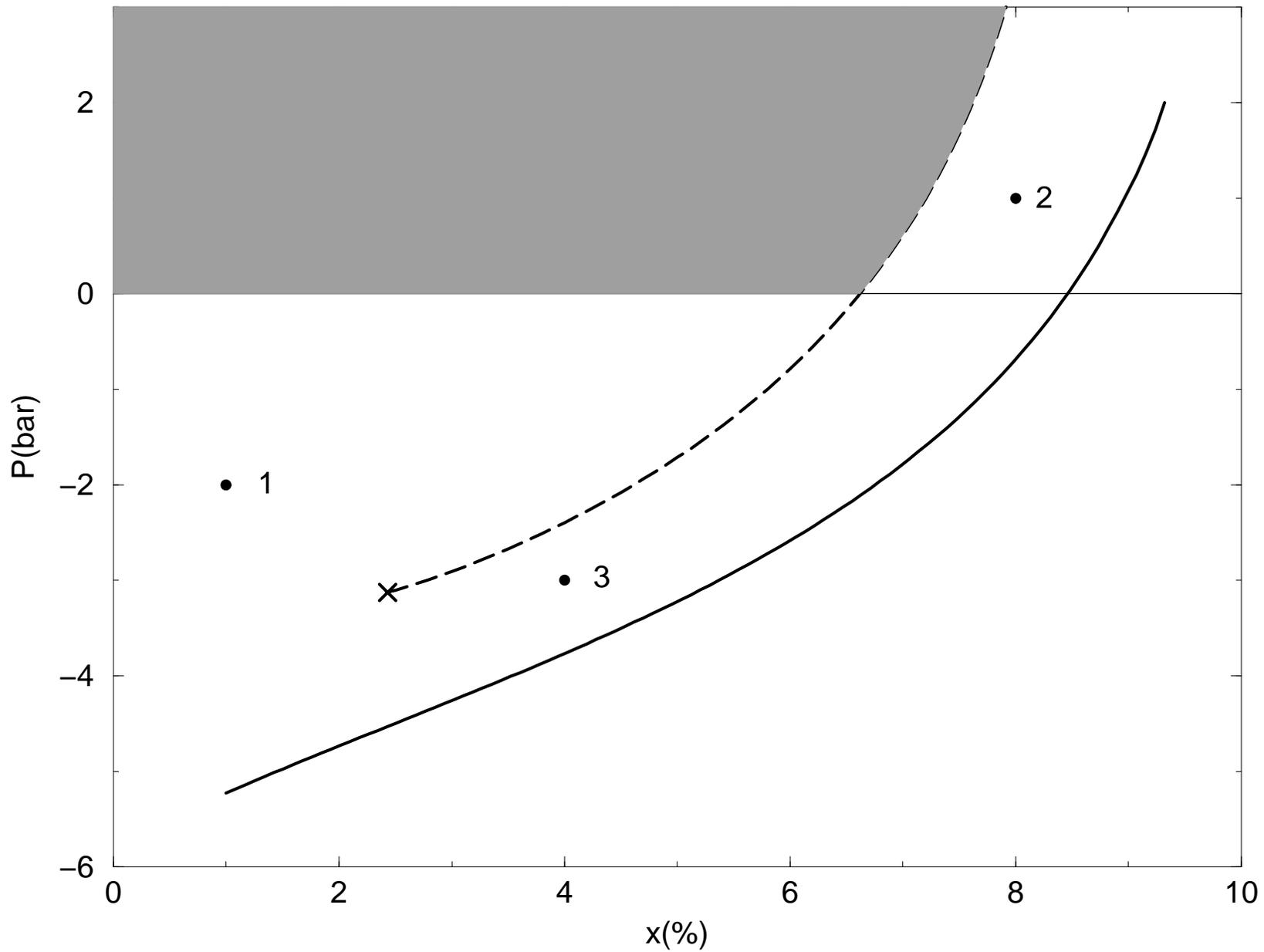
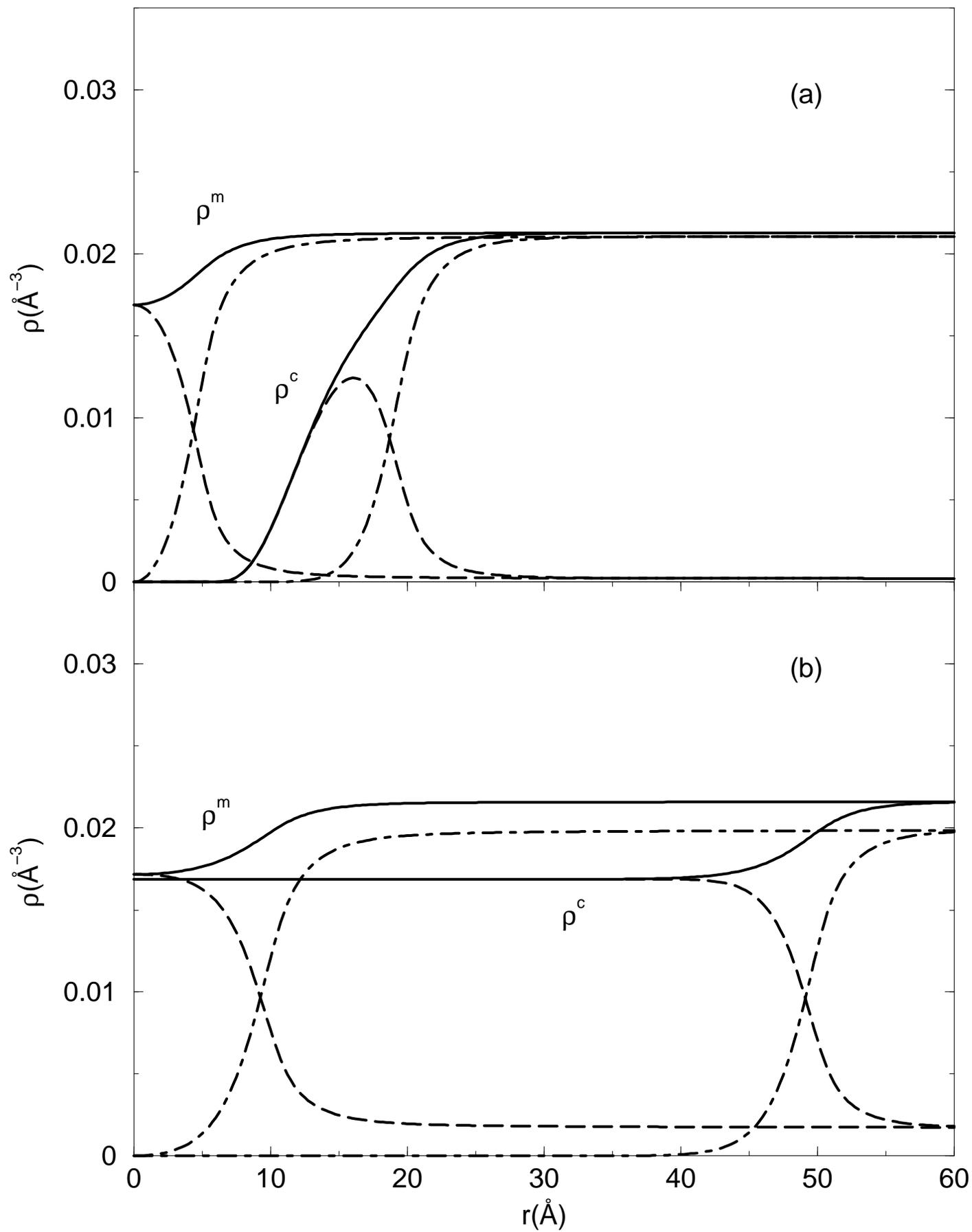
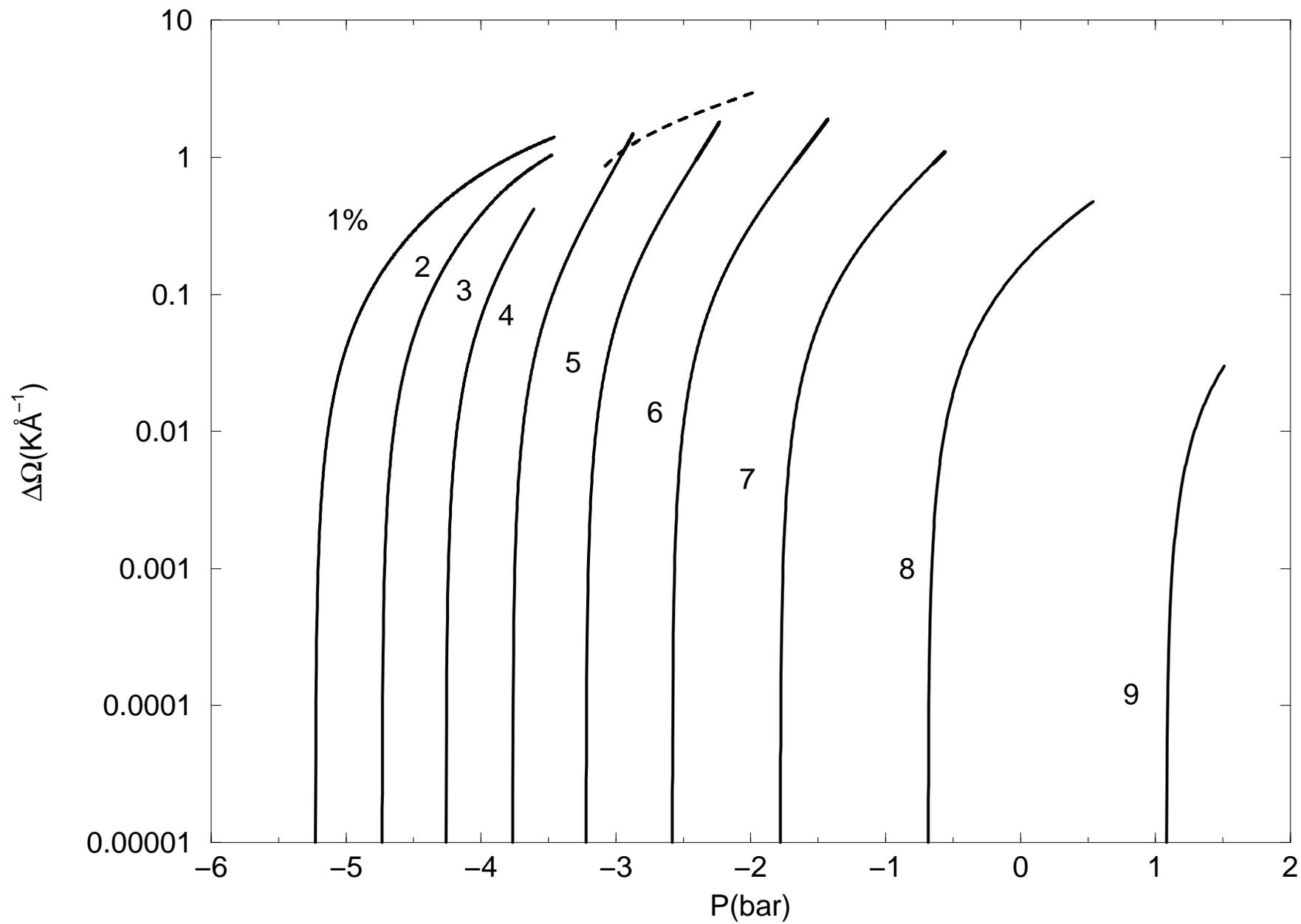
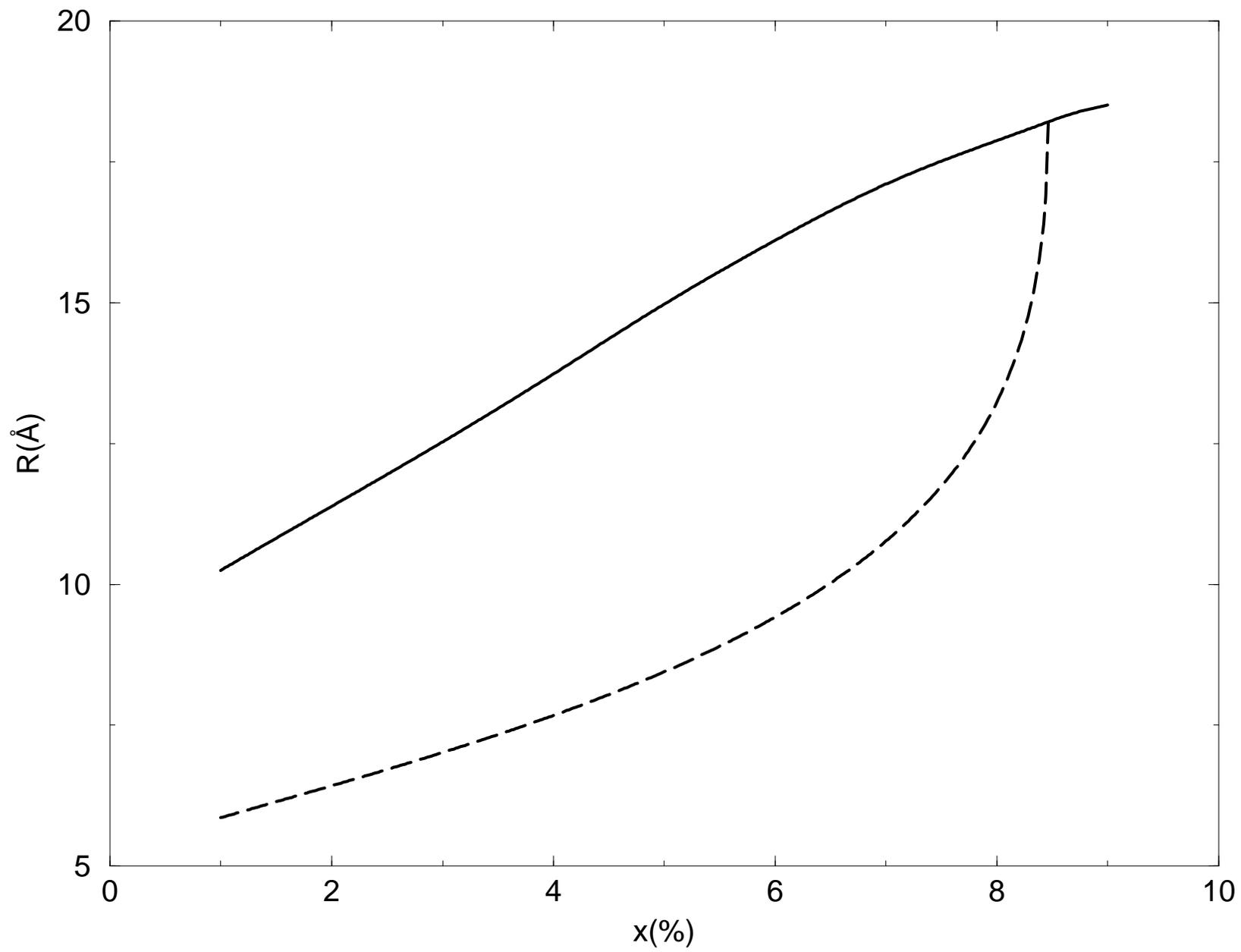
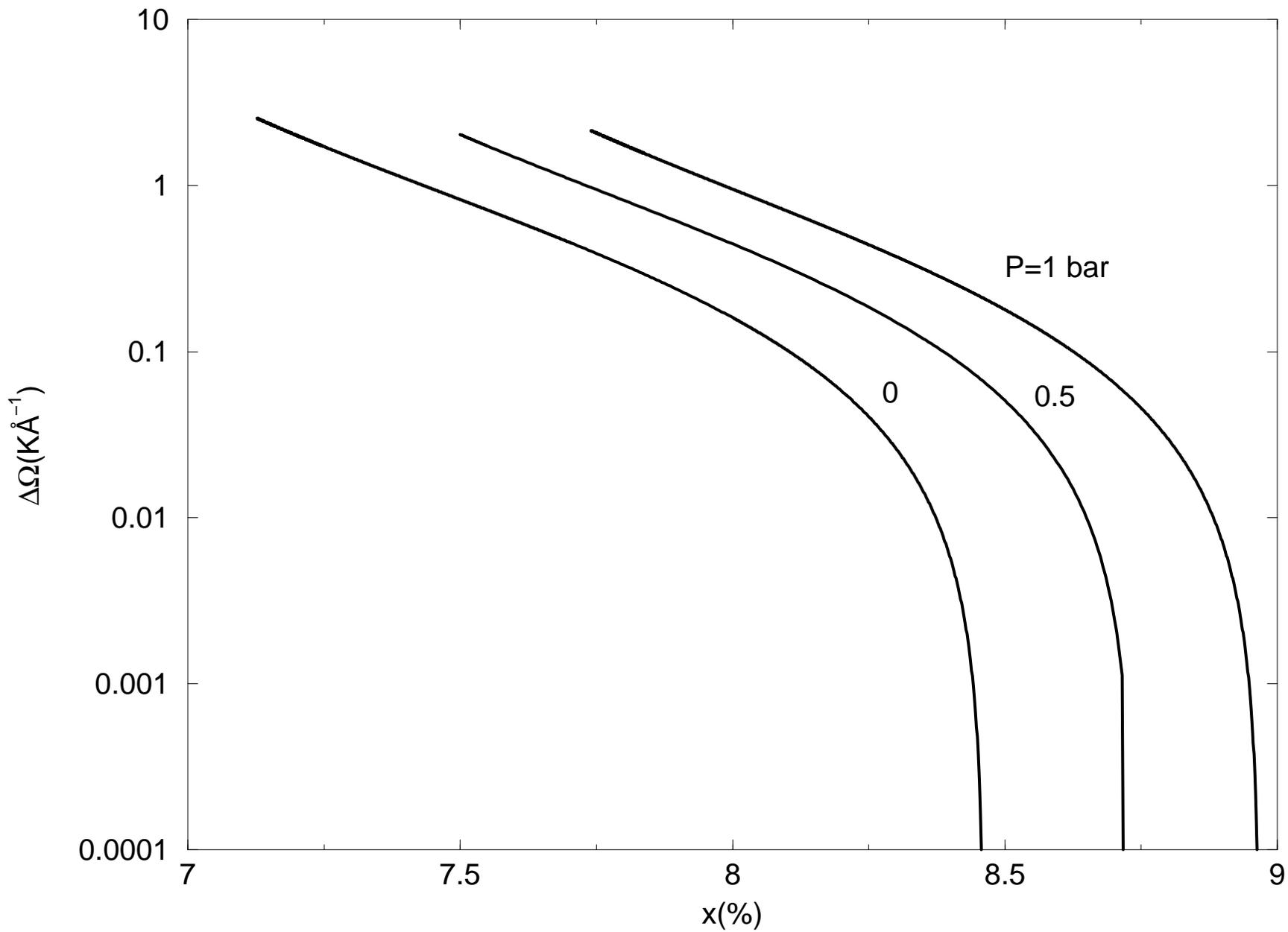
\end{document}